\documentclass[letterpaper,10pt,journal,twoside]{IEEEtran}
\usepackage{cite}

\ifCLASSINFOpdf
   \usepackage[pdftex]{graphicx}
   \DeclareGraphicsExtensions{.pdf,.jpeg,.png}
\else
   \usepackage[dvips]{graphicx}
   \DeclareGraphicsExtensions{.eps}
\fi
\usepackage{amsmath}
\usepackage{amssymb}  

\usepackage{algorithmicx}
\usepackage{algpseudocode}
\usepackage{algorithm}

\usepackage{hyperref}
\usepackage{fancyhdr}

\newtheorem{definition}{Definition}
\newtheorem{theorem}{Theorem}
\newtheorem{lemma}{Lemma}

\newtheorem{proposition}{Proposition}
\newtheorem{problem}{Problem}

\newtheorem{remark}{Remark}

\makeatletter
\newcommand\fs@betterruled{%
	\def\@fs@cfont{\bfseries}\let\@fs@capt\floatc@ruled
	\def\@fs@pre{\vspace*{5pt}\hrule height.8pt depth0pt \kern2pt}%
	\def\@fs@post{\kern2pt\hrule\relax}%
	\def\@fs@mid{\kern2pt\hrule\kern2pt}%
	\let\@fs@iftopcapt\iftrue}
\floatstyle{betterruled}
\restylefloat{algorithm}
\makeatother

\begin{document}
%
\title{
	On-Line Permissive Supervisory Control of\\Discrete Event Systems for scLTL Specifications
}
%
%
%

\author{Ami~Sakakibara,~\IEEEmembership{Student~Member,~IEEE,}
        and~Toshimitsu~Ushio,~\IEEEmembership{Member,~IEEE}
\thanks{
	A.~Sakakibara and T.~Ushio are with 
	the Department of Systems Innovation, 
	Graduate School of Engineering Science, Osaka University, 
	1-3 Machikaneyama, Toyonaka, Osaka, 560-8531, Japan
	(e-mail: {\small sakakibara@hopf.sys.es.osaka-u.ac.jp}; {\small ushio@sys.es.osaka-u.ac.jp}).
}
\thanks{
	This work was supported by JST ERATO Grant Number JPMJER1603, Japan, and 
	JSPS KAKENHI Grant Number JP19J13487, Japan. 
}
}

\maketitle
\thispagestyle{fancy}
\renewcommand{\headrulewidth}{0pt}
\chead{\footnotesize{Citation information: DOI 10.1109/LCSYS.2020.2971029, IEEE Control Systems Letters}}
\cfoot{\scriptsize{(c) 2020 IEEE.  Personal use of this material is permitted. See http://www.ieee.org/publications\_standards/publications/rights/index.html for more information.}
}

\begin{abstract}
We propose an on-line supervisory control scheme for discrete event systems (DESs), 
where a control specification is described by a fragment of linear temporal logic. 
On the product automaton of the DES and an acceptor for the specification, 
we define a ranking function that returns the minimum number of steps required to reach an accepting state from each state. 
In addition, we introduce a permissiveness function that indicates a time-varying permissive level. 
At each step during the on-line control scheme, the supervisor refers to the permissiveness function as well as the ranking function in order to guarantee the control specification while handling the tradeoff between its permissiveness and acceptance of the specification.
The proposed scheme is demonstrated in a surveillance problem for a mobile robot.
\end{abstract}


\ifCLASSOPTIONpeerreview
\begin{center} \bfseries EDICS Category: 3-BBND \end{center}
\fi
%
\IEEEpeerreviewmaketitle

\section{Introduction}
%
%
%
%

\IEEEPARstart{T}{he} main idea underlying supervisory control of discrete event systems (DESs) \cite{Cassandras2008}, 
initiated by Ramadge and Wohnam \cite{Ramadge1987}, 
is to restrict behavior of the system by appropriately enabling some controllable events. 
A supervisor issues its control action so that the sequences generated hereafter will be kept within a predefined specification language.  
One difficulty in this framework is that control requirements are typically given by formal languages, i.e., subsets of event sequences generated by the system.  
Practically, we need to translate desired behavior of the system into formal languages, which is a hard task.
For this reason, linear temporal logic (LTL) \cite{Baier2008} is paid much attention to as a formal specification language for control problems, thanks to its rich expressiveness. 
Controller synthesis problems with temporal logic specifications are widely studied for many types of system models \cite{Belta2017,Ding2014a,Tumova2016,Jiang2006,Sakakibara2018b}.
It is practically acceptable to restrict the specification language to a subclass of LTL like syntactically co-safe LTL (scLTL), for which synthesis problems can be solved in much less complexity than for the case of the general LTL \cite{Kupferman2001}.
We consider a supervisory control problem of a DES under an scLTL constraint, which describes more general characterization of system properties than the conventional marking-based modeling.

Synthesis of supervisors turns out to be computationally hard when the controlled system is large or has much complex aspects.
Some researchers have overcome the difficulty by designing supervisors on-line \cite{Chung1992,Chung1993,Chung1994,Hadj-Alouane1996,Heymann1994,Prosser1998,Grigorov2006}.
Chung et al.~ proposed a method to generate limited lookahead trees on-the-fly instead of constructing a complete supervisor \cite{Chung1992}.
Heymann and Lin took another approach for on-line supervisory control under partial observation, where the on-line supervisor modifies appropriate control actions precomputed in the case of full observation \cite{Heymann1994}.

We propose an on-line control scheme leveraging a \textit{ranking function} that enables us to find desirable behavior with respect to the scLTL specification. 
The concept of ranking functions is like that of Lyapunov functions, which play a great role in determining control strategies.
The key idea is that, if the rank decreases along a trajectory, we regard it as good.
Ranking functions are useful in solving games played on a graph and reachability analysis of automata, which sometimes give solutions to LTL-related problems. 
The function proposed in \cite{Ding2014a} for receding horizon control under LTL constraints is based on a similar idea and captures distances to accepting states with respect to a given LTL specification. 
However, their approach is not directly applicable to our problem because they do not consider uncontrollable events.

In our approach, the ranking function is defined on the product automaton of the DES and the specification automaton. 
The rank decreases if an accepting state of the product automaton is being approached, i.e., the scLTL specification is more likely to be satisfied than the previous time step. 
From a view point of supervisory control theory, on the other hand, it is desirable for the supervisor to enable as many events as possible. 
Thus, we additionally introduce a \textit{permissiveness function} that indicates a time-varying permissiveness level.
At each step of the on-line control process, the supervisor checks the current permissiveness level. 
If the permissive level is high, the supervisor may enable events that do not necessarily lead to achievement of the specification. 
By referring to the ranking function and the permissiveness function, which quantify distances to acceptance and the permissiveness level, respectively, the supervisor explicitly handles the tradeoff between the acceptance of the scLTL specification and its permissiveness.

\IEEEpubidadjcol

\section{Preliminaries}
\label{sec:preliminaries}

For a set $T$, we denote by $ |T| $ its cardinality. 
For an element $ t \in T $ and a subset $ T' \subseteq T $, we define an indicator function $ I_{T'}:T \to \{0,1\} $ as $ I_{T'}(t) = 1 $ if $ t \in T' $ and $ I_{T'}(t) = 0 $ otherwise. 

$T^*$ (resp., $T^\omega$) is a set of finite (resp., infinite) sequences over $T$. 
For a finite or infinite sequence $\tau$ over $T$, i.e., $\tau \in T^*$ or $\tau \in T^\omega$, let $\tau[j]$ be the $(j+1)$-th element of $\tau$.
For any $ j,k \in \mathbb{N} $ with $ j \leq k $, we denote by $ \tau[j \ldots k] $ the sequence $ \tau[j]\tau[j+1]\ldots\tau[k] $. 
We write $ \tau' \preceq \tau $ if $ \tau' $ is a prefix of $ \tau $. 
$ \|\tau\| $ stands for the length of a finite sequence $ \tau \in T^* $. 

\subsection{System Model}
\subsubsection{Transition Systems} \label{sec:TS}
A (deterministic) transition system is a 4-tuple
$ \mathcal{T} = (X, \Sigma, \delta, x_0) $, 
where 
$X$ is the set of states, 
$\Sigma$ is the set of events, a partial function $\delta: X \times \Sigma \to X$ is the transition function, and
$x_0 \in X$ is the initial state. 
We write $\delta(x,\sigma)!$ if a transition from $x$ with $\sigma$ is defined. 
For each $x \in X$, let $\Sigma(x)=\{ \sigma \in \Sigma: \delta(x, \sigma)! \}$ \cite{Hadj-Alouane1996} and $ \mathsf{suc}_\mathcal{T}(x) = \{x' \in X: \exists \sigma \in \Sigma(x), x'=\delta(x,\sigma) \} $.
We denote by a triple $ (x,\sigma,x') $ a transition from $ x \in X $ to $ x' = \delta(x,\sigma) $ for some $ \sigma \in \Sigma(x) $. 
Moreover, the transition function is extended to a sequence of inputs: for $\sigma \in \Sigma$ and $s \in \Sigma^*$, 
$ \delta(x,\varepsilon) = x $ and $\delta(x,s\sigma)= \delta(\delta(x,s),\sigma)$.

Let $ \mathcal{L}(\mathcal{T}) = \{ s \in \Sigma^*: \delta(x_0,s)! \} $ be the set of all finite event sequences generated by $ \mathcal{T} $. 
An infinite sequence $ \rho \in X(\Sigma X)^\omega $ is called a {\it run} if, for any $j \in \mathbb{N}$, $\rho[2(j+1)] = \delta(\rho[2j], \rho[2j+1])$.
A finite sequence $ h \in X(\Sigma X)^* $ is called a {\it history} if $ h \in X $ or, for $ h $ with $ \|h\|\geq 3 $, $  h[2(j+1)] = \delta (h[2j], h[2j+1]) $ for any $j \in \{ 0, 1, \ldots, \frac{\|h\|-3}{2} \}$. 
The set of all runs (resp., histories) starting from the initial state $ x_0 $ is defined as $ \mathsf{Runs}(\mathcal{T}) $ (resp., $ \mathsf{His}(\mathcal{T}) $).

\subsubsection{Discrete Event Systems}
We model a discrete event system (DES) by a labeled transition system \cite{Jiang2006} 
$ G=((X,$ $\Sigma,\delta,x_0), AP, L) $, 
where the first tuple $(X,\Sigma,\delta,$ $x_0)=: \mathcal{T}_G$ is a deterministic transition system, 
$AP$ is the set of atomic propositions, namely, the set of simple known statements that are either true or false \cite{Baier2008}, and 
$L: X \to 2^{AP}$ is the labeling function.
The event set is partitioned into disjoint subsets $\Sigma = \Sigma_c \cup \Sigma_u$, where $\Sigma_c$ (resp., $\Sigma_u$) is the set of controllable (resp., uncontrollable) events.
We define $ \Sigma_c(x) = \Sigma(x) \cap \Sigma_c $ and $ \Sigma_u(x) = \Sigma(x) \cap \Sigma_u $ for each $ x \in X $. 
$ p \in L(x) $ means that an atomic proposition $ p \in AP $ holds at a state $ x \in X $, i.e., $ p $ is true at $ x $.
$G$ is said to be finite if $X$, $\Sigma$, and $AP$ are all finite. 
We define $ \mathcal{L}(G) = \mathcal{L}(\mathcal{T}_G) $, $ \mathsf{Runs}(G) = \mathsf{Runs}(\mathcal{T}_G) $, and $ \mathsf{His}(G) = \mathsf{His}(\mathcal{T}_G) $.
For simplicity, we write $ \mathsf{suc}_G $ for $ \mathsf{suc}_{\mathcal{T}_G} $.

For the set $AP$, a {\it word} is a finite or infinite sequence of a subset $\nu \in 2^{AP}$ of atomic propositions, called a {\it letter}.
Each run in $\mathsf{Runs}(G)$ generates an infinite word over $2^{AP}$, obtained by the labeling function.
We extend the labeling function as follows: for run $\rho$,
$L(\rho)= L(\rho[0]) L(\rho[2]) \ldots $.
The extension for histories is defined similarly.

\subsection{Syntactically Co-Safe Linear Temporal Logic}

Linear temporal logic (LTL) is useful to describe qualitative control specifications. 
We focus on syntactically co-safe LTL (scLTL), a subclass of LTL.
Formally, an scLTL formula $\varphi$ over the set $ AP $ of atomic propositions is defined as
\[
\varphi \! ::= \! \mathrm{true} \ | \  a \ |\ \neg a \ |\ \varphi_1 \land \varphi_2 \ | \ \varphi_1 \lor \varphi_2 \ | \ \bigcirc \varphi \ | \ \varphi_1 \! \mathbf{U} \varphi_2,  
\] 
where $ a \in AP $, $\varphi, \varphi_1, \varphi_2$ are scLTL formulas. 
In addition, we use a temporal operator $\Diamond$ defined as 
$ \Diamond\varphi := \mathrm{true} \mathbf{U} \varphi $.

The semantics of scLTL is defined over an infinite word \cite{Baier2008}. 
Intuitively, $ \bigcirc \varphi $ is true if $ \varphi $ holds on the word from the {\em next} step and 
$ \varphi_1 \mathbf{U} \varphi_2 $ if $ \varphi_1 $ keeps to be satisfied {\em until} $ \varphi_2 $ turns true at some step. 
For an scLTL formula $ \varphi $ over $ AP $ and an infinite word $ w \in (2^{AP})^\omega $, we write $ w \models \varphi $ if $ w $ satisfies $ \varphi $. 
For a DES $ G $ and an scLTL formula $ \varphi $, we say $ G $ satisfies $ \varphi $, denoted by $ G \models \varphi $, if $ L(\rho) \models \varphi $ for all $ \rho \in \mathsf{Runs}(G) $. 

Although scLTL formulas are evaluated over infinite words, it is known that we only need to check whether an input word has a \textit{good prefix} of the formula. 
Any scLTL formula can be translated into a corresponding deterministic finite automaton (DFA), which is an acceptor for the good prefixes \cite{Kupferman2001}.
For an scLTL formula $ \varphi $, let $ A_\varphi = ((X_A, \Sigma_A, \delta_A, x_{A,0}), F_A) $ be its corresponding DFA, where $ (X_A, \Sigma_A, \delta_A, x_{A,0}) $ is a deterministic transition system with $ \Sigma_A=2^{AP} $ and $ F_A \subseteq X_A $ is the set of accepting states. 
For any $ w \in (2^{AP})^\omega $, we have 
\begin{equation}
\label{eq:acc_scLTL}
w \!\models \!\varphi \!\! \iff \! \exists w'\! \!\in \!(2^{AP})^*\!, w' \preceq w \land \delta_A(x_{A,0},w') \!\in \!F_A. 
\end{equation}

\section{Formulation}
\label{sec:formulation}

We formulate a supervisory control problem with an scLTL specification. 
A controller, called a \textit{supervisor}, enables some controllable events at each state \cite{Ramadge1987}. 
For each state $ x \in X $, we define $ \Gamma(x) = \{ \gamma \subseteq \Sigma(x): \Sigma_u (x) \subseteq \gamma \} $,
where 
$ \gamma \in \Gamma (x) $ is called a \textit{control pattern} at $ x $. 
Let $ \Gamma=\bigcup_{x \in X} \Gamma(x)$ be the set of all control patterns. 
The supervisor observes the history from the initial state to the current state and determines a control pattern. 
Formally, we define a \textit{path-based supervisor} as a mapping $ \mathcal{S}: \mathcal{L}(G) \to \Gamma $ such that, for each $ s \in \mathcal{L}(G) $, $ \mathcal{S}(s) \in \Gamma(\delta(x_0,s)) $. 

\begin{definition}[Supervised behavior]
	Let $ \mathcal{S} $ be a path-based supervisor for a DES $ G = ((X, \Sigma, \delta, x_0), AP, L) $. 
	The closed-loop behavior of the DES $ G $ under the control by $ \mathcal{S} $, denoted by $ \mathcal{S}/G $, is given by 
	\begin{align*}
	&\varepsilon \in \mathcal{L}(\mathcal{S}/G), \text{ and } \\
	&\forall s \in \mathcal{L}(\mathcal{S}/G), s\sigma \in \mathcal{L}(\mathcal{S}/G) 
	\iff s \sigma \in \mathcal{L}(G)  \land \sigma \in \mathcal{S}(s).  \\
	&\mathsf{Runs}(\mathcal{S}/G) = \{ x_0 \sigma_1 x_1 \ldots \in \mathsf{Runs}(G): \sigma_1 \in \mathcal{S}(\varepsilon) \\ &\qquad \qquad \qquad \qquad \land \forall j \geq 1, \sigma_{j+1} \in \mathcal{S}(\sigma_1 \sigma_2 \ldots \sigma_{j}) \}. 
	\end{align*}
\end{definition}

\begin{definition}[Finite-state supervisor]
	A \textit{finite-state supervisor} for a DES $ G = ((X, \Sigma, \delta, x_0), AP, L) $ is a tuple $ S=((M, \Sigma, \mu_M, m_0),\mu_C) $, where 
	$ (M, \Sigma, \mu_M, m_0) $ is a deterministic transition system with the finite state set $ M $ 
	and a feedback function $ \mu_C: M \times \mathbb{N} \to \Gamma $. 	
\end{definition}

If the supervisor is in state $ m \in M $ at step $ k \in \mathbb{N} $, the supervisor assigns $ \mu_C(m,k) $ as a control pattern to the DES.
\begin{definition}[Supervisor realization]
	\label{def:realization}
	Let $ \mathcal{S} $ be a path-based supervisor for a DES $ G = ((X, \Sigma, \delta, x_0), AP, L) $. 
	A finite-state supervisor $ S=((M, \Sigma, \mu_M, m_0),\mu_C) $ \textit{realizes} $ \mathcal{S} $ if, 
	for all $ s \in \mathcal{L}(G) $, 
	$ \mathcal{S}(s) = \mu_C \big( \mu_M(m_0, s), \|s\| \big) $. 
\end{definition}

\begin{problem}
	\label{prob:main}
	Given a finite DES $ G=((X,\Sigma,\delta,x_0),$  $ AP, L) $ and an scLTL formula $ \varphi $ over $ AP $, 
	synthesize a supervisor $ \mathcal{S} $ such that $ \mathcal{S}/G \models \varphi $.
\end{problem}

\section{Ranking Function for Product Automaton}
To solve Problem \ref{prob:main}, we design an on-line supervisor, which dynamically computes a control pattern at each time step. 
In our control scheme, we first execute the preprocessing off-line and then move on to the on-line control stage, where the supervisor stops control after detecting a history corresponding to a good prefix of the scLTL specification.
In this section, we explain the off-line computation shown in Algorithm \ref{alg:pm}, which outputs a ranking function.

The specification scLTL formula $ \varphi $ is converted into an equivalent DFA $ A_\varphi $.
Then, we compute the product automaton $ P $ of the DES $ G = ((X,\Sigma,\delta,x_0),AP,L) $ and the DFA $ A_\varphi = ((X_A,2^{AP},\delta_A,x_{A,0}),F_A) $, defined as follows. 
\begin{equation*}
\label{eq:product}
P = G \otimes A_\varphi = ((X_P,\Sigma_P,\delta_P,x_{P,0}),F_P),
\end{equation*}
where $ (X_P,\Sigma_P,\delta_P,x_{P,0}) $ is a deterministic transition system with 
$ X_P = X \times X_A $, 
$ \Sigma_P = \Sigma $, 
$ \delta_P:X_P \times \Sigma_P \to X_P $,  
$ x_{P,0} = (x_0,\delta_A(x_{A,0},L(x_0))) $, and 
$ F_P = X \times F_A $. 
For each $ x = (x_G,x_A) \in X_P $ and $ \sigma \in \Sigma_P $, 
$ \delta_P(x,\sigma) = \big(\delta(x_G,\sigma), \delta_A(x_A, L(\delta(x_G,\sigma))\big) $ and we define $ J_G(x) = x_G $ and $ J_A(x) = x_A $.
Note that $ \mathcal{L}(P) = \mathcal{L}(G) $. 

Since the product automaton captures the behavior of the DES and the DFA at the same time, our goal turns out to reach an accepting state of the product automaton. 
For that purpose, we introduce a ranking function that measures a distance to accepting states. 
Intuitively, the rank of a state $ x $ represents the minimum number of steps required to reach an accepting state from $ x $ under some control. 
We define $ \alpha = |X_P|-|F_P|+1 $ as the upper bound of the ranking for the product automaton $ P $. 
If $ x $ is ranked as $ \alpha $, then it is impossible to force the product automaton to reach an accepting state from $ x $, i.e., either 
1) $ x $ is among a strongly connected component from which no accepting state is reachable, or 
2) it is inevitable to go to such a nonaccepting sink from $ x $ because of uncontrollable events. 
Formally, a ranking function is defined as follows. 
\begin{definition}
	\label{def:rank_func}
	Let $ P=((X_P,\Sigma_P,\delta_P,x_{P,0}),F_P) $ be a product automaton and $ \alpha = |X_P|-|F_P|+1 $. 
	A function $ \xi: X_P \to \mathbb{N} $ is a \textit{ranking function} for $ P $ if both of the following conditions hold for any $ x \in X_P $.
	\begin{align*}
	1) &\ \xi(x) = 0 \iff x \in F_P; \\
	2) &\ \xi(x) =
	\begin{cases}
	\alpha & \text{if } \Sigma_P(x) = \emptyset, \\
	\min \big\{
	\displaystyle \min_{\sigma \in \Sigma_{P,c}(x)} \xi (\delta_P(x,\sigma)) &\!\!\!\!\!\!+ I_{X_P\setminus F_P}(x),\alpha \big\}  \\ &\text{if } \Sigma_{P,u}(x) = \emptyset, \\
	\min \big\{	
	\displaystyle \max_{\sigma \in \Sigma_{P,u}(x)} \xi (\delta_P(x,\sigma)) &\!\!\!\!\!\!+ I_{X_P\setminus F_P}(x),\alpha \big\}\\& \text{otherwise.} 
	\end{cases}
	\end{align*}
\end{definition}

The ranking function is obtained by Algorithm \ref{alg:pm}, the correctness of which can be proved similarly to \cite{Bernet2002}.
As initialization, we set $ \xi(x) = 0 $ for each $ x \in X_P $ and $ \alpha = |X_P|-|F_P|+1 $. 
Then, we go on to update the values of $ \xi $ by using functions $ \hat{\xi}: X_P \to \mathbb{N} $ 
and $ \mathsf{up}_\alpha : \mathbb{N} \times X_P \to \mathbb{N} $, defined as follows.
For each $ x \in X_P $, 
\begin{align*}
&\hat{\xi}(x) =
\begin{cases}
\alpha & \text{if } \Sigma_P(x) = \emptyset, \\
\displaystyle
\min_{\sigma \in \Sigma_{P,c}(x)} \xi( \delta_P(x,\sigma) ) & \text{if }\Sigma_{P,u}(x) = \emptyset, \\
\displaystyle
\max_{\sigma \in \Sigma_{P,u}(x)} \xi(\delta_P(x,\sigma)) & \text{otherwise. } 
\end{cases}
\end{align*}
For any $ r \in \mathbb{N} $ and $ x \in X_P $, 
\begin{align*}
&\mathsf{up}_\alpha (r, x) 
= 
\begin{cases}
r+1 & \text{ if } x \notin F_P \land r < \alpha ,  \\
r & \text{ otherwise. }
\end{cases}
\end{align*}
To sum up, the current rank is incremented if the current state is not accepting with at least one uncontrollable event defined and the rank has not hit the upper bound;
otherwise the rank does not change.

\begin{algorithm}[!tb]
	\centering
	\caption{Off-line computation} \label{alg:pm}
	\begin{algorithmic}[1]
		\Require{A DES $ G = ((X, \Sigma, \delta, x_0),L, AP) $ and an scLTL formula $ \varphi $ over $ AP $. 
		}
		\Ensure{A product automaton $ P = ((X_P,\Sigma_P,\delta_P,x_{P,0}),$ $F_P) $ and a ranking function $ \xi: X_P \to \mathbb{N} $.}
		\State Construct a DFA $ A_\varphi $ from the scLTL formula $ \varphi $.
		\State Compose the DES $ G $ and the DFA $ A_\varphi $ into the product automaton $ P$.
		\State Compute a ranking function $ \xi $ for $ P $: 
		\State $ \alpha \gets |X_P|-|F_P|+1 $ \label{line:pm_init_start}
		\ForAll{$ x \in X_P $} 
		\State $ \xi(x) \gets 0 $
		\EndFor \label{line:pm_init_end}
		\While{$ \exists x \in X_P $ s.t. $ \xi(x)  < \mathsf{up}_\alpha ( \hat{\xi}(x) ,x) $} \label{line:pm_update_start}
		\State $ \xi (x) \gets \mathsf{up}_\alpha (\hat{\xi}(x),x)$
		\EndWhile \label{line:pm_update_end}
	\end{algorithmic}
\end{algorithm}

\begin{proposition}
	\label{prop:suc_lower_rank}
	For any $ x \in X_P $, if $ 0 < \xi(x) <\alpha $ then 
	\begin{align*}
	\mathsf{suc}_P(x) \cap \{ x' \in X_P: \xi(x) > \xi(x') \} \neq \emptyset .
	\end{align*}
\end{proposition}

\begin{IEEEproof}
	Let $ x \in X_P $ be an arbitrary state such that $ 0 < \xi(x) < \alpha $. 
	Since $ \xi(x) > 0 $, $ x $ is an nonaccepting state, i.e., $ x \notin F_P $ and thus $ I_{X_P\setminus F_P}(x) = 1 $. 
	Suppose that, for all $ x' \in \mathsf{suc}_P(x) $, $ \xi(x) \leq \xi(x') $.
	By the second condition of Definition \ref{def:rank_func}, if $ \Sigma_{P,u}(x) = \emptyset $, then we have 
	\begin{align*}
	\xi(x) \!= \!\!\!\min_{\sigma \in \Sigma_{P,c}(x)} \!\!\xi (\delta_P(x,\sigma)) \!+\! I_{X_P\setminus F_P}(x) 
	\geq \xi (x) \!+\! 1 > \xi(x).
	\end{align*}
	By contradiction, we conclude that there exists $ x' \in \mathsf{suc}_P(x) $ satisfying $ \xi(x) > \xi(x') $. 
	For the case of $ \Sigma_{P,u}(x) \neq \emptyset $, we follow a similar discussion. 
\end{IEEEproof}

Proposition \ref{prop:suc_lower_rank} ensures that a lower-ranked successor always exists.
Then, by taking such a successor at each step, the product automaton eventually reaches an accepting state. 

\begin{proposition}
	\label{prop:xi_valid}
	For any $ x \in X_P $, it holds that 
	\[ \xi(x) < \alpha \implies \exists s \in \Sigma_P^*, \delta_P(x,s) \in F_P. \]
\end{proposition}

\begin{IEEEproof}
	Let $ x \in X_P $ such that $ \xi(x) < \alpha $. 
	From Proposition \ref{prop:suc_lower_rank}, there exists $ s \in \Sigma_P^* $ along which the rank decreases, i.e., 
	$ \xi(x) > \xi(\delta_P(x,s[0])) > \xi(\delta_P(x,s[0]s[1])) > \ldots > \xi(\delta_P(x,s)) = 0 $.
	By the first condition of Definition \ref{def:rank_func}, $ \delta_P(x,s) \in F_P $. 
\end{IEEEproof}

\begin{proposition}
	\label{prop:uc_included}
	For any $ x \in X_P \setminus F_P $, if $ \xi(x) < \alpha $ then 
	\[ \Sigma_{P,u}(x) \subseteq \{ \sigma \in \Sigma_P(x): \xi(x) > \xi(\delta_P(x,\sigma)) \}. \]
\end{proposition}

\begin{IEEEproof}
	Let $ x \in X_P \setminus F_P $ such that $ \xi(x) < \alpha $ and $ \Sigma_{P,u}(x) \neq \emptyset $.
	By the second condition of Definition \ref{def:rank_func},  
	\[
	\forall \sigma \in \Sigma_{P,u}(x),\ \xi(x) \geq \min \big\{ \displaystyle \xi (\delta_P(x,\sigma)) + I_{X_P\setminus F_P}(x),\alpha \big\}.
	\]
	Thus, we have $ \xi(x) \geq \xi(\delta_P(x,\sigma)) + 1 > \xi(\delta_P(x,\sigma)) $ for any $ \sigma \in \Sigma_{P,u}(x) $, which completes the proof.
\end{IEEEproof}

Proposition \ref{prop:uc_included} says that the rank always decreases along a transition triggered by an uncontrollable event.
By Proposition \ref{prop:uc_included}, we obtain the following proposition. 
\begin{proposition}
	\label{prop:u_seq}
	For any $ x \in X_P $ with $ \xi(x) < \alpha $,
	\[ \forall u \in \Sigma_{P,u}^*, \ \delta_P(x,u)! \implies \xi(\delta_P(x,u)) < \alpha.  \]
\end{proposition}

By Propositions \ref{prop:xi_valid} and \ref{prop:u_seq}, from any state $ x $ with $ \xi(x) < \alpha $, it is possible for the product automaton to eventually reach an accepting state regardless of the occurrences of uncontrollable events hereafter. 
We characterize transitions of the product automaton with respect to the ranking function.
Let $ (x,\sigma,x')\in X_P \times \Sigma_P \times X_P $ be a transition defined in the product automaton.
\begin{itemize}
	\item $ (x,\sigma,x') $ is \textit{legal} with respect to $ \xi $ if $ \xi(x) > \xi(x') $. 
	\item $ (x,\sigma,x') $ is \textit{neutral} with respect to $ \xi $ if $ \xi(x) \!\leq\! \xi(x') \!<\! \alpha $. 
	\item $ (x,\sigma,x') $ is \textit{illegal} with respect to $ \xi $ if $ \xi(x') = \alpha $. 
\end{itemize}

It is possible to lead the product automaton to reach an accepting state 
if we always choose legal transitions. 
On the other hand, visiting higher-ranked states is also acceptable to some extent if their ranks do not hit the upper bound $ \alpha $. 
Indeed, we are likely to obtain more \textit{permissive} supervisors if we allow not only legal transitions but also neutral ones to be enabled. 
Permissiveness is one of the most important concepts in supervisory control theory, where 
we often aim to design a supervisor that enables as many events as possible. 

It should be noticed that, however, the number of occurrences of neutral transitions must be limited.
Since otherwise, livelock may occur, i.e., the product automaton may stay within states $ x $ with $ 0 < \xi(x) < \alpha  $ while it always holds the possibility of reaching an accepting state but actually suspends going there.  
That is, we have a tradeoff between permissiveness of the supervisor and achievement of the specification. 
To take the tradeoff into consideration, we introduce a criterion for how many neutral transitions we allow to be enabled, which plays a key role in the on-line control scheme explained in the next section.

\section{On-Line Permissive Supervisory Control}
Our goal is to design a supervisor that determines its control action on-line, being aware of a time-varying permissiveness level. 
We start from introducing a function that quantifies the permissiveness level.

\begin{definition}
	\label{def:energy}
	A \textit{permissiveness function} is a function $ \eta:\mathbb{N} \to \mathbb{R} $ that satisfies the following three conditions. 
	\begin{enumerate}
		\item $ \eta(0) \leq \alpha $;
		\item $ \eta(k) \geq \eta(k+1) $ for any $ k \in \mathbb{N} $;
		\item $ \eta(\bar{k}) = 0 $ for some $ \bar{k} \in \mathbb{N} $.
	\end{enumerate}
\end{definition}
That is, the permissiveness level decreases as time goes by and will eventually hit the lower bound $ 0 $. 

The on-line supervisor, denoted by $ \hat{\mathcal{S}} $, is realized by a finite-state supervisor $ \hat{S}=(\mathcal{T}_P, \mathsf{online}) $, where for each $ x \in X_P $ and $ k \in \mathbb{N} $, 
\begin{equation}
\label{eq:online}
\mathsf{online}(x,k) \!
= \! \big\{ \sigma \!\in\! \Sigma_P\!:\! \xi(\delta_P(x,\sigma)) \!<\!\max\{ \xi(x), \eta(k) \}\! \big\}.
\end{equation}

The on-line supervisor refers to the current rank $ \xi(x) $ and the current permissiveness level $ \eta(k) $ to take into consideration the tradeoff between permissiveness and acceptance of the specification. 
More precisely, the higher permissiveness level we have, the more neutral transitions we allow to be enabled. 
Since the permissiveness level decreases with the elapse of time, the function $ \mathsf{online} $ returns less events as time goes by. 
In the following, we show properties of $ \mathsf{online} $. 
\begin{lemma}
	\label{lem:controlpattern}
	For any $ x \in X_P $ and $ k \in \mathbb{N} $, $ \mathsf{online}(x,k) \in \Gamma(J_G(x)) $.
\end{lemma}

\begin{IEEEproof}
	Let $ x \in X_P $ and $ k \in \mathbb{N} $ be an arbitrary state and a nonnegative number, respectively. 
	We prove that $ \Sigma_u(x_G) \subseteq \mathsf{online}(x,k) $, where $ x_G = J_G(x) $. 
	Note that $ \Sigma_P(x)=\Sigma(x_G) $ and $ \Sigma_{P,u}(x) = \Sigma_u(x_G) $.
	From Eq.~(\ref{eq:online}), $ \sigma \in \mathsf{online}(x,k) $ means that 
	$ \xi(\delta_P(x,\sigma)) < \xi(x) $ or $ \xi(\delta_P(x,\sigma)) < \eta(k) $. 
	Thus, from Proposition \ref{prop:uc_included}, we conclude that $ \Sigma_u(x_G) = \Sigma_{P,u}(x) \subseteq \{ \sigma \in \Sigma_P: \xi(\delta_P(x,\sigma)) < \xi(x)  \} \subseteq \mathsf{online}(x,k) $. 
\end{IEEEproof}

\begin{lemma}
	For any state $ x \in X_P $ and step $ k \in \mathbb{N} $, if $ 0 < \xi(x) < \alpha $, then $ \mathsf{online}(x,k) \neq \emptyset $. 
\end{lemma}

\begin{IEEEproof}
	For any $ x \in X_P $ and $ k \in \mathbb{N} $, we have 
	\begin{eqnarray*}
		\label{eq:online_inclusion}
		\mathsf{online}(x,k) \supseteq 
		\{ \sigma \in \Sigma_P: \xi(x) > \xi(\delta_P(x,\sigma)) \} \neq \emptyset, 
	\end{eqnarray*}
	where the nonemptiness is guaranteed by Proposition \ref{prop:suc_lower_rank}.
\end{IEEEproof}

\begin{proposition}
	\label{prop:eventually_empty}
	For any $ x \in X_P $ with $ 0 < \xi(x) < \alpha $, there exists $ k \in \mathbb{N} $ such that, for any $ l \geq k $,  
	\[ \{ \sigma \in \Sigma_P: \xi(\delta_P(x,\sigma)) < \eta(l) \} = \emptyset. \]
\end{proposition}

\begin{IEEEproof}
	By the second and third conditions of Definition \ref{def:energy}, there exists $ \bar{k} \in \mathbb{N} $ such that, for all $ \bar{l} \geq \bar{k} $, $ \eta(\bar{l})=0 $.
	Since the ranking function $ \xi $ returns nonnegative values, $ \{ \sigma \in \Sigma_P: \xi(\delta_P(x,\sigma)) < 0 \} = \emptyset $ for any $ x \in X_P $.
\end{IEEEproof}

\begin{lemma}
	The on-line supervisor $ \hat{\mathcal{S}} $ is a path-based supervisor for $ G $. 
\end{lemma}

\begin{IEEEproof}
	It holds by Definition \ref{def:realization} that, 
	for any $ s \in \mathcal{L}(G) $, 
	$ \hat{\mathcal{S}}(s) = \mathsf{online}(\delta_P(x_{P,0},s),\|s\|) $. 
	By Lemma \ref{lem:controlpattern}, we have $ \hat{\mathcal{S}}(s) \in \Gamma(J_G(\delta_P(x_{P,0},s))) = \Gamma(\delta(x_0,s)) $.
\end{IEEEproof}

\subsubsection*{On-Line Control Scheme}
If $ \xi(x_{P,0}) < \alpha $, the on-line control scheme starts under initialization with $ m = x_{P,0} $ and $ k = 0 $. 
At each step $ k $ after the event string $ s \in \mathcal{L}(G) $ with $ \|s\|=k $, the on-line supervisor $ \hat{\mathcal{S}} $ computes a control pattern by the function $ \mathsf{online} $ if $ \xi(m) > 0 $. 
Then, the supervisor observes the event $ \sigma_{k+1} $ that the DES $ G $ has executed after given the control pattern $ \mathsf{online}(\delta_P(x_{P,0},s),k) $. 
According to the observation, the supervisor updates its state  and time step to $ m=\delta_P(x_{P,0},s\sigma_{k+1}) $ and $ k+1 $, respectively, and then goes on to determine the next control action. 
If $ \xi(m) = 0 $, then the supervisor stops the control process. 

\begin{lemma}
	\label{lem:neutral_limited}
	Assume that $ \xi(x_{P,0}) < \alpha $.
	Under the control by the on-line supervisor $ \hat{\mathcal{S}} $, 
	neutral transitions of the product automaton occur only finitely often.
\end{lemma}

\begin{IEEEproof}
	Suppose that neutral transitions occur infinitely often.
	That is, there exists $ s \in \Sigma^\omega $ such that 
	\begin{equation}
	\label{eq:ass}
	\begin{array}{l}
	\!\!\!\!\forall k \in  \mathbb{N}, \exists l \geq k, \\
	\xi(\delta_P(x_{P,0}, s[0\ldots l])) \leq \xi( \delta_P(x_{P,0}, s[0\ldots l+1]) ).
	\end{array}
	\end{equation}
	Let $ \sigma_{l+1}\!:=\!s[l\!+\!1] $ for some $ l\! \in\! \mathbb{N} $ that satisfies the above inequality and $ x_l:= \delta_P(x_{P,0},s[0\ldots l]) $.
	Since $ \sigma_{l+1}\! \in \!\hat{\mathcal{S}}(s[0\dots l]) = \mathsf{online}(x_l,l) $, we have  
	$ \xi( \delta_P(x_l, \sigma_{l+1}) ) \!<\! \eta(l) $. 
	Eq.~(\ref{eq:ass}) says that there exist infinitely many $ l \in \mathbb{N} $ such that $ \xi( \delta_P(x_l, \sigma_{l+1}) )\! < \!\eta(l) $, which contradicts Proposition \ref{prop:eventually_empty}.
\end{IEEEproof}

\begin{lemma}
	\label{lem:online_accepting}
	Assume that $ \xi(x_{P,0}) < \alpha $.
	An accepting state of the product automaton is eventually reached 
	under the control by the on-line supervisor $ \hat{\mathcal{S}} $.
\end{lemma}

\begin{IEEEproof}
	Recall that $ \eta(0) \leq \alpha $ and that $ \eta $ is nonincreasing, as mentioned in Definition \ref{def:energy}.
	Since the on-line supervisor $ \hat{\mathcal{S}} $ never allows illegal transitions, 
	we have, for any $ s \in \mathcal{L}(\hat{\mathcal{S}}/G) $, $ \xi(\delta_P(x_{P,0},s)) < \alpha $. 
	By Proposition \ref{prop:xi_valid}, then, it is always possible to lead the product automaton to an accepting state by some appropriate event sequence.
	From Lemma \ref{lem:neutral_limited}, while the on-line computation is running, the supervisor $ \hat{\mathcal{S}} $ observes neutral transitions only finitely often. 
	Let $ \bar{k} \in \mathbb{N} $ be the step index of the last occurrence of a neutral transition. 
	Then, for any $ \bar{l} \geq \bar{k} $ and $ x \in X_P $, $ \mathsf{online}(x,\bar{\ell}) = \{ \sigma \in \Sigma_P: \xi(\delta_P(x,\sigma)) < \xi(x) \} $.
	In other words, the supervisor $ \hat{\mathcal{S}} $ chooses only legal transitions after time step $ \bar{k} $. 
	Since the rank always decreases during each legal transition, eventually a state ranked as $ 0 $, namely, an accepting state is reached. 
\end{IEEEproof}

\begin{theorem}
	$ \hat{\mathcal{S}}/G \models \varphi $ if $ \xi(x_{P,0}) < \alpha $. 
\end{theorem}

\begin{IEEEproof}
	From Lemma \ref{lem:online_accepting}, the on-line supervisor forces the DES $ G $ to generate event sequences with which the product automaton eventually reaches an accepting state. 
	Note that, when an accepting state of the product automaton is reached, the corresponding word is accepted by the DFA $ A_\varphi $. 
	By Eq.~(\ref{eq:acc_scLTL}), any run that has the corresponding history as a prefix satisfies the scLTL formula $ \varphi $.
\end{IEEEproof}

\begin{remark}
	Our control problem is based on $ \omega $-languages, for which there does not exist a maximally permissive finite-state supervisor \cite{Ehlers2016}. 
	This is why we do not apply the traditional method \cite{Cassandras2008} but take a different approach, 
	considering a time-varying permissiveness level. 
	One may synthesize a maximally permissive supervisor computed by regarding the DFA $ A_\varphi $ as an acceptor for a $ * $-language specification.
	However, the resulting supervisor does not always satisfy the scLTL specification in the supervised DES due to the existence of livelock situations, which must be avoided in our setting. 
	When our method is applied to supervisory control problems discussed in a language-based manner, on the other hand, 
	the proposed on-line supervisor is not maximally permissive. 
	Despite the difference, the proposed on-line supervisor may determine a maximally permissive control pattern that includes not only all legal transitions but also all neutral ones if the permissiveness level is sufficiently close to $ \alpha $. 
\end{remark}

\section{Illustrative Example}
\subsubsection{Scenario}

In this section, we apply the proposed on-line supervisory control scheme to a surveillance problem, 
where a single mobile robot moves around the environment consisting of six rooms and collects data by attached sensors.

The DES $ G $ is given by the synchronous product \cite{Cassandras2008} of two DESs $ G_{pos} $ and $ G_{task} $, which correspond to the location of the robot and the sensing task module, respectively. 
The graph structures of their transition system are depicted in Fig.~\ref{fig:modules}. 
$ G_{pos} $ has the state set $ X_{pos} = \{ x_0, x_1, \ldots, x_5 \}  $ and 
the event set $ \Sigma_{move} = \{ \sigma_0,\sigma_1,\ldots,\sigma_5 \} $, where by $ \sigma_i $ the robot goes to or stays at Room $ i $. 
We assume that all events in $ \Sigma_{move} $ are controllable. 
The robot is initially located in Room 0 and thus we set state $ x_0 $ as the initial state of $ G_{pos} $. 
On the other hand, $ G_{task} $ has the state set $ X_{task} = \{y_0,y_1,y_2\} $ and the event set $ \Sigma_{task} \cup \Sigma_{move} $, 
where $ \Sigma_{task} = \{ \sigma_{start}, \sigma_{stop}, \sigma_{comp}, \sigma_{idle} \} $ with $ \sigma_{comp} $ and $ \sigma_{idle} $ being uncontrollable. 
The robot moves from room to room by events in $ \Sigma_{move} $, which are shared with $ G_{pos} $.
It starts collecting data by $ \sigma_{start} $ and $ G_{task} $ moves to the state $ y_2 $ by $ \sigma_{comp} $ if the sensing task has been completed. 
From state $ y_2 $, the uncontrollable event $ \sigma_{idle} $ lets the robot return back to the initial state $ y_0 $, where another command is acceptable.
Sensing can be interrupted by the controllable event $ \sigma_{stop} $ even before its completion.
The composite DES $ G $ has 18 states and 40 transitions. 

Let $ AP = AP_{pos} $ $\cup AP_{task} $ be the set of atomic propositions for $ G $, where $ AP_{pos} = \{ p_0,p_1,\ldots,p_5 \} $ and $ AP_{task} = \{ q_s \} $. 
Each atomic proposition in $ AP_{pos} $ represents the current location of the robot, while the atomic proposition $ q_s $ specifies the situation where the robot finishes collecting data. 
The labeling function $ L $ of the whole model $ G $ is given by the following rules. 
Let $ (x, y) \in X_{pos} \times X_{task} $ be an arbitrary state of the composite DES. 
\begin{itemize}
	\item $ p_i \in L( x, y ) \iff x = x_i $ for each $ i = 0,1,\ldots,5 $; 
	\item $ q_s \in L( x, y ) \iff y = y_2 $.
\end{itemize}

It is required that the robot should complete collecting data at Room 4 and then go to Room 3 to do so, before returning back to Room 0. 
The control specification is formally given by an scLTL formula 
\begin{equation}
\label{eq:spec}
\!
\begin{array}{l@{\ }l}
\varphi = &  \bigcirc \big( \neg p_0 \mathbf{U} (p_3 \land q_s)\big)\land \bigcirc \big( \neg p_0 \mathbf{U} (p_4 \land q_s)\big)  \\ 
&\ \ \land \bigcirc \big( \neg (p_3 \land q_s) \mathbf{U} (p_4 \land q_s)\big) \land \bigcirc \Diamond (p_0 \land \neg q_s) .
\end{array}
\!\!\!\!
\end{equation}

\subsubsection{Results}

\begin{figure}[tbp]
	\centering
	\includegraphics[bb=0 0 355 110, width=0.99\linewidth]{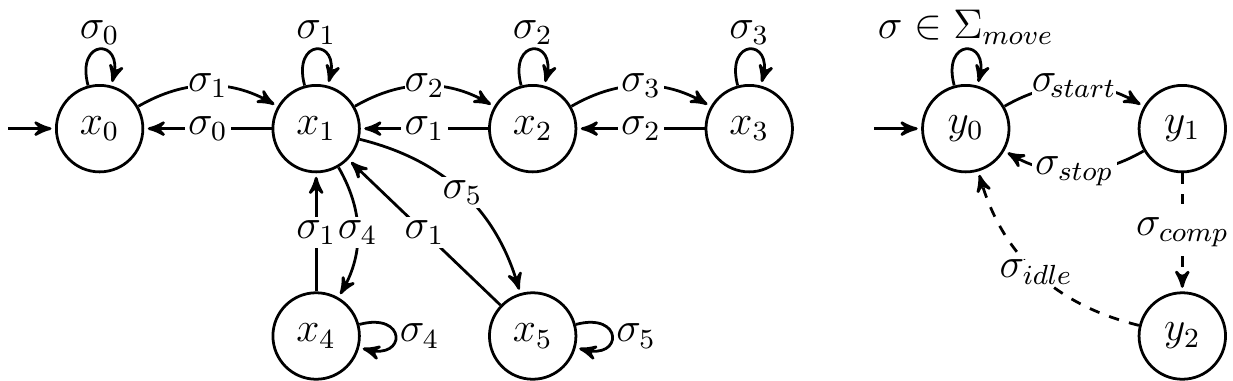}
	\caption{
		The graph structures of $ G_{pos} $ (left) and $ G_{task} $ (right).
		Solid and dashed arrows represent transitions triggered by controllable and uncontrollable events, respectively. 
	}
	\label{fig:modules}
	\vspace*{10pt}
	\includegraphics[bb=0 2 361 105, width=0.99\linewidth]{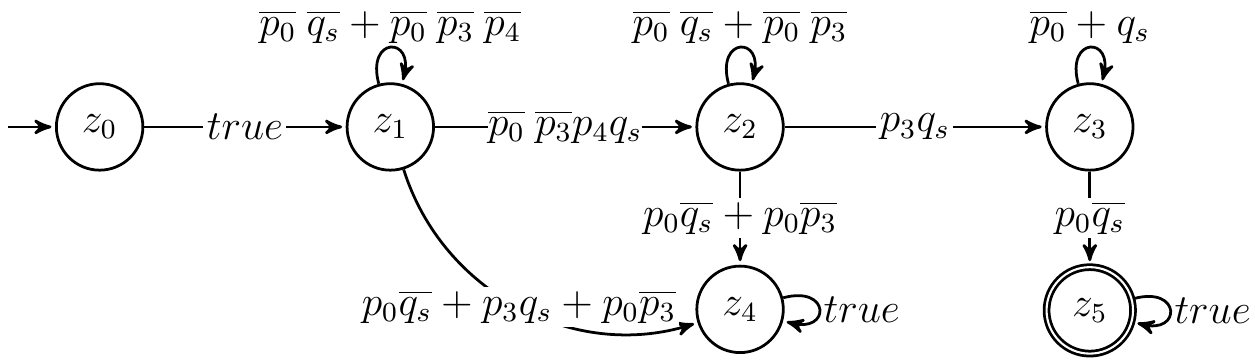}
	\caption{
		The DFA $ A_\varphi $ that accepts good prefixes for $ \varphi $, where $ F_A = \{z_5\} $.
		Transitions labeled with $ true $ are triggered by any letter $ \nu \in 2^{AP} $. 
		For $ p,q \in AP $, transition labels of the form $ \overline{p} $, $ pq $, and $ p+q $ represent letters $ \nu \in 2^{AP} $ satisfying $ p\notin \nu $, $ p \in \nu \land q \in \nu $, and $ p \in \nu \lor q \in \nu $, respectively.
		We only show transitions triggered by $ \nu \in \{ \nu' \in 2^{AP}: |\nu' \cap AP_{pos}| = 1 \} $ because no state in $ X_{pos} $ is associated with more than two position labels. 
	}
	\label{fig:dfa}
\end{figure}

In the off-line computation, we use Spot\footnote{https://spot.lrde.epita.fr/} to convert the scLTL formula into a corresponding DFA. 
From the control specification formula $ \varphi $ given by Eq.~(\ref{eq:spec}), we obtain a DFA shown in Fig.~\ref{fig:dfa}.
The accessible part of the product automaton has 79 states and 182 transitions.
Since the rank of the initial state $ \xi(x_{P,0}) $ is 14, which is smaller than $ \alpha = 91 $\footnote{We compute $ \alpha=|X_P|-|F_P|+1 $ from the size of the whole product automaton, including the unreachable part.}, 
we move on to the on-line control scheme.

In the on-line computation, we use a permissiveness function of the form $ \eta(k) = \max \{ ak + b, 0 \} $ with $ a<0 $ and $ b\leq \alpha $. 
At each step of the simulation, an event that the DES executes is selected randomly from the control pattern given by the on-line supervisor. 
Shown in Fig.~\ref{fig:trajectory_plotter} are the traces of the rank of states in a trajectory generated by $ \hat{\mathcal{S}}/G $, where the parameters of the permissiveness function are set to (i) $ (a,b) = (-0.5,20) $ and (ii) $ (a,b) = (-0.5,30) $. 
In both cases, the supervisor finally reaches an accepting state, with the rank of zero, which means that a good prefix of the scLTL specification is detected. 
We can also see that neutral transitions are enabled until the permissiveness level declines to be lower than the rank of the current state.

We additionally examine the difference among controlled behavior derived from different permissiveness functions. 
More precisely, we change the value of the parameter $ a $ of the permissiveness function while the parameter $ b $ is fixed as $ 30 $ and compare the results. 
We execute the on-line control scheme 1000 times for each configuration of the permissiveness function and then calculate 
(i) the average number of steps taken to reach an accepting state and 
(ii) the average size of control patterns given by the on-line supervisor. 
The summary of the simulation results is shown in Fig.~\ref{fig:compare}. 
Notice that, the larger the absolute value of $ a $ is, the sooner the supervisor reaches an accepting state and the smaller control patterns it computes.

\section{Conclusions}
We propose an on-line supervisory control scheme for DESs to achieve a control specification given by scLTL formulas. 
A ranking function computed off-line helps the supervisor find good successors in light of the scLTL specification. 
In the on-line computation, the supervisor refers to a permissiveness function, which indicates a time-varying permissiveness level, together with the ranking function and improves its permissiveness if possible. 
Our approach is also practically applicable to, e.g., surveillance problems where 
the system must surely complete the mission but it is not required to terminate very soon. 
It is future work to extend the proposed scheme to supervisory control under partial observation or under general LTL constraints. 

{
\renewcommand{\arraystretch}{0.05}
\begin{figure}[tbp]
	\centering 
	\hspace*{-7.5pt}
	\begin{tabular}{c}
		\begin{minipage}{0.49\hsize}
			\centering
			\includegraphics[width=\linewidth]{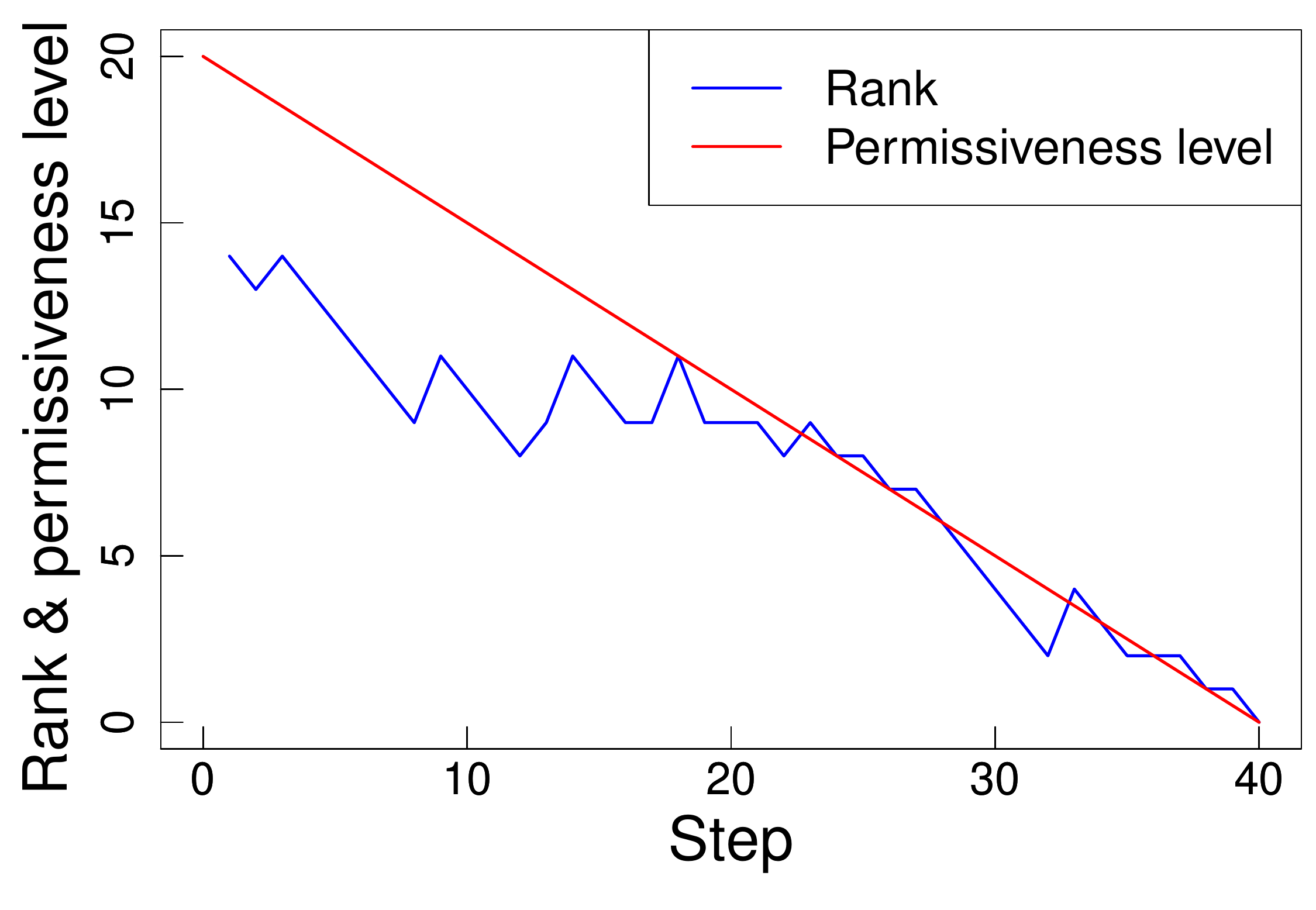}
		\end{minipage}	
		\begin{minipage}{0.49\hsize}
			\includegraphics[width=\linewidth]{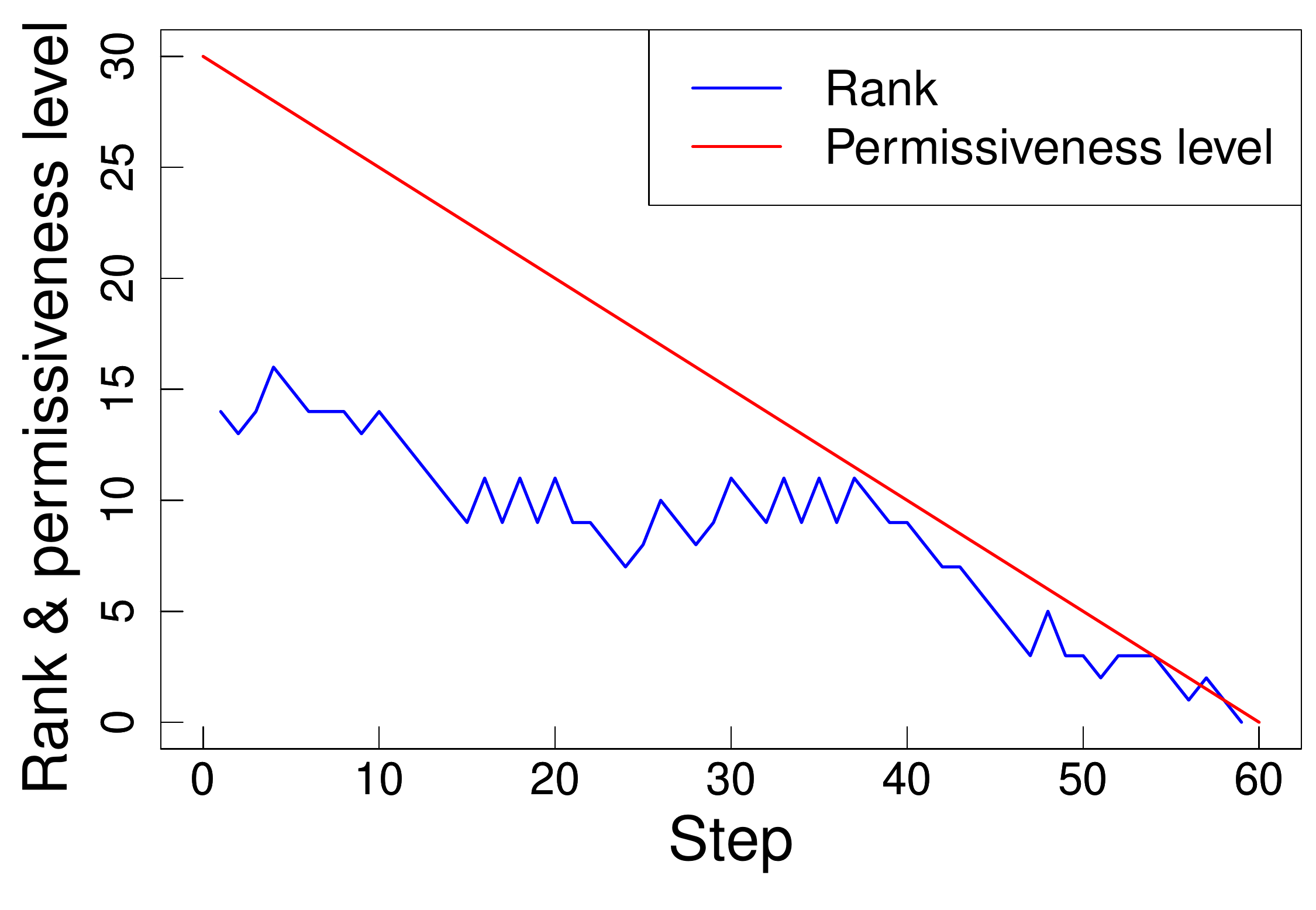}
		\end{minipage}
		\\
		\begin{minipage}{0.49\hsize}
			\centering
			{\scriptsize (i) $ (a,b) = (-0.5, 20) $.}
		\end{minipage}
		
		\begin{minipage}{0.49\hsize}
			\centering
			{\scriptsize (ii) $ (a,b) = (-0.5, 30) $.}
		\end{minipage}
	\end{tabular}
	\caption{Plots of the rank $ \xi(x) $ at each state $ x $ visited in a sample trajectory, until getting to $ 0 $, and permissiveness functions with different parameters. }
	\label{fig:trajectory_plotter}
	\centering \vspace*{10pt} 
	\hspace*{-7.5pt}
	\begin{tabular}{c}
		\begin{minipage}{0.49\hsize}
			\centering
			\includegraphics[width=\linewidth]{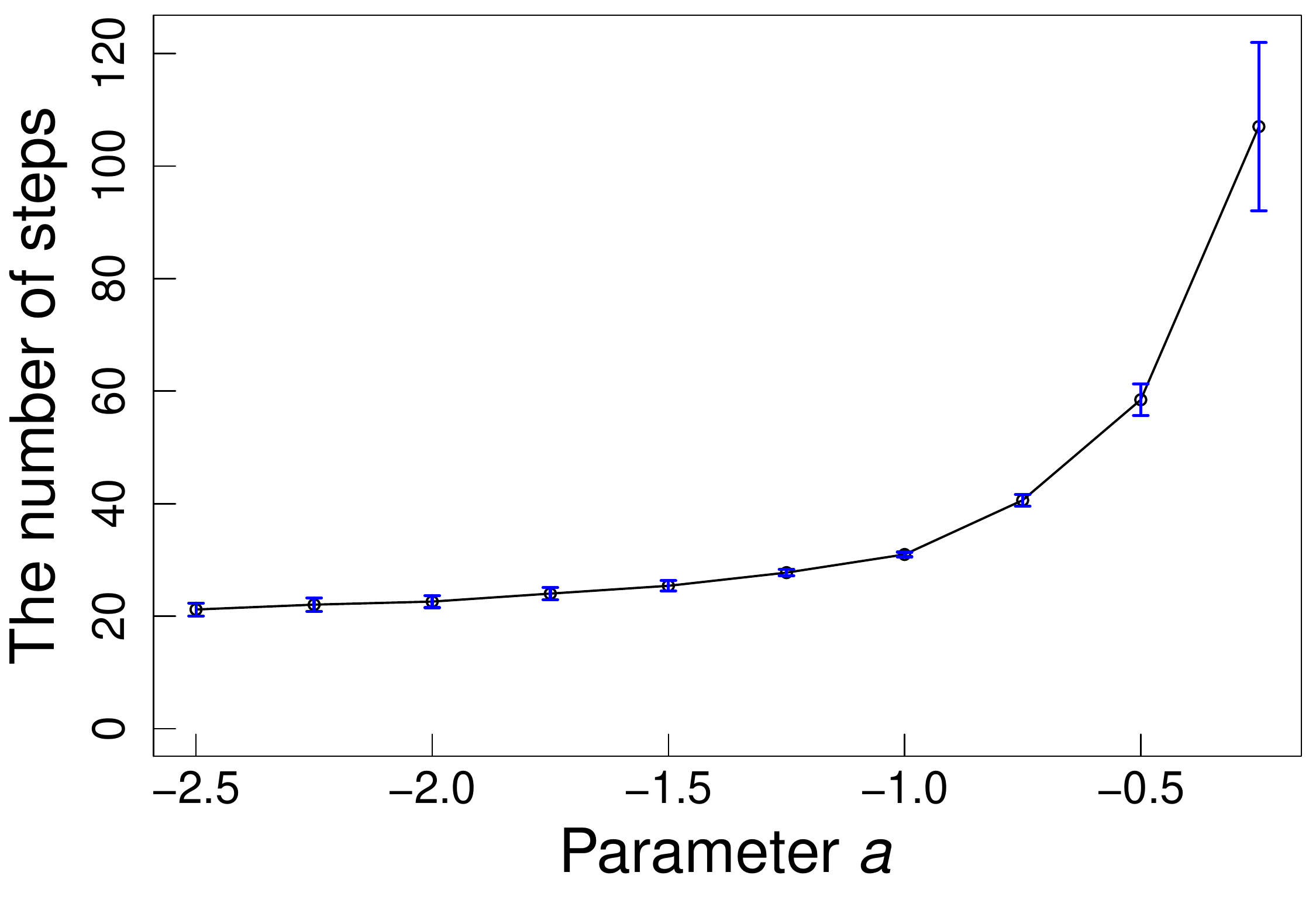}
		\end{minipage}	
		
		\begin{minipage}{0.49\hsize}
			\includegraphics[width=\linewidth]{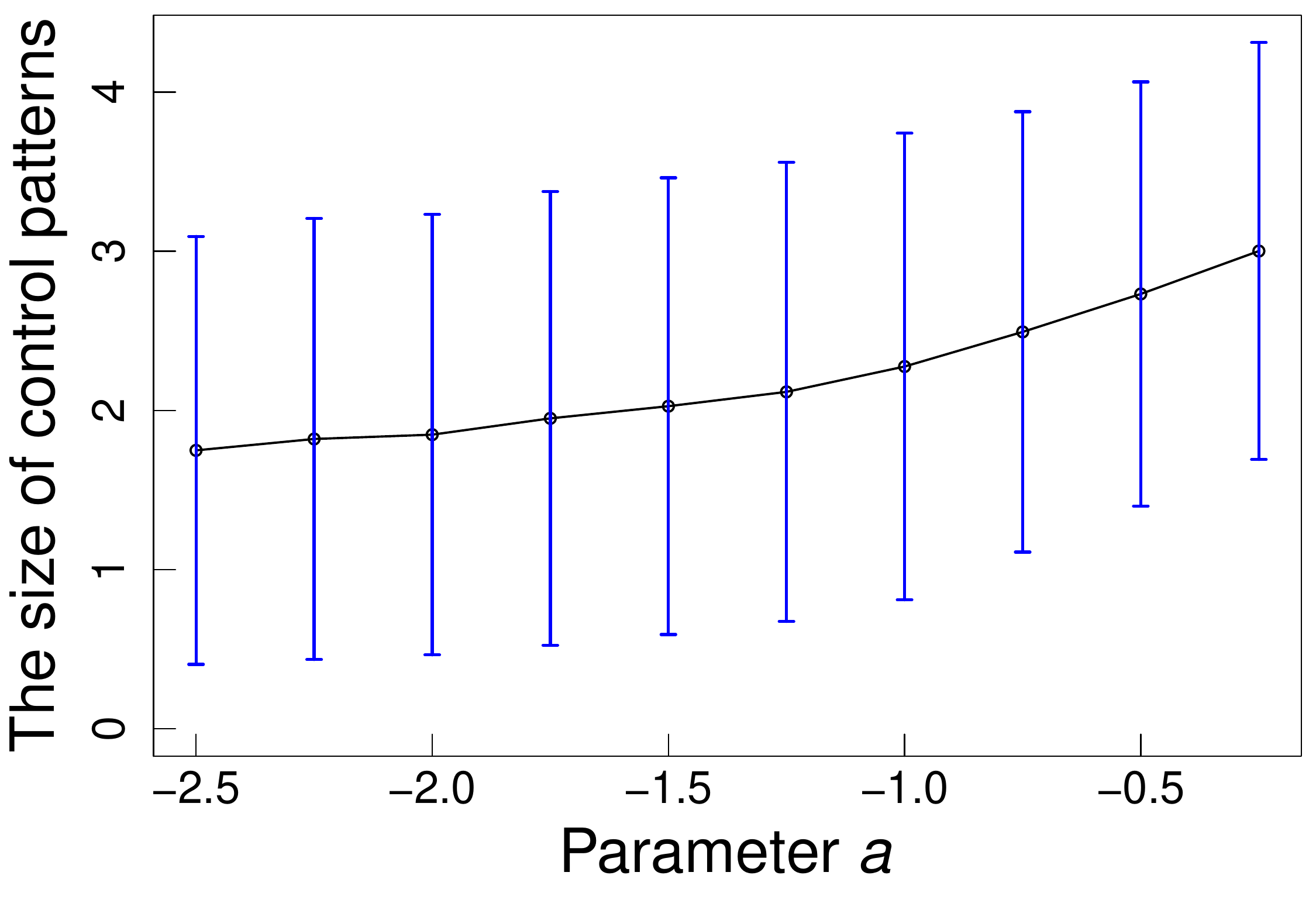}
		\end{minipage}
		\\
		\begin{minipage}{0.49\hsize}
			\centering 
			{\scriptsize (i) } 
		\end{minipage}
		
		\begin{minipage}{0.49\hsize}
			\centering 
			{\scriptsize (ii) } 
		\end{minipage}
	\end{tabular}
	\caption{The relation between the permissiveness function of the form $ \eta(k) = \max\{ak+30,0\} $. 
		Plots of 
		(i) the average number of steps taken to reach an accepting state and 
		(ii) the average size of control patterns computed by the on-line supervisor, 
		where the error bars show the standard deviations.
	}
	\label{fig:compare}
\end{figure}
}

\appendices


\section*{Acknowledgment}
The authors would like to thank Prof.~Ichiro Hasuo for his informative comments on the terminology.

\ifCLASSOPTIONcaptionsoff
  \newpage
\fi



\bibliographystyle{IEEEtran}
\bibliography{bibdata}
\end{document}